\documentclass[sigconf,9pt,authorversion=True]{acmart}

\usepackage{tikz}
\usepackage{graphicx}

\makeatletter
\renewcommand\@formatdoi[1]{\ignorespaces}
\makeatother

\usepackage{subcaption}
\setcopyright{acmcopyright}
\copyrightyear{2024}
\acmYear{2024}
\acmDOI{10.1145/1122445.1122456}

\acmConference[N2Women '24]{N2Women '24: Network Networking Women Professional Development}{June 03, 2024}{Tokyo, Japan}
\acmBooktitle{N2Women '24: Network Networking Women Professional Development, June 03, 2024, Tokyo, Japan}

\usepackage[flushleft]{threeparttable}
\begin{document}

\title{Detecting abnormal heart sound using mobile phones and on-device IConNet}

\author{Linh Vu}
\email{linh.vu@monash.edu}
\orcid{0000-0002-6313-0860}
\affiliation{
  \institution{Monash University}
  \country{Malaysia}
}

\author{Thu Tran}
\email{ndttran.2019@phdcs.smu.edu.sg}
\orcid{0009-0003-2854-3525}
\affiliation{
  \institution{Singapore Management University}
  \country{Singapore}
}

\begin{abstract}
Given the global prevalence of cardiovascular diseases, there is a pressing need for easily accessible early screening methods. Typically, this requires medical practitioners to investigate heart auscultations for irregular sounds, followed by echocardiography and electrocardiography tests. To democratize early diagnosis, we present a user-friendly solution for abnormal heart sound detection, utilizing mobile phones and a lightweight neural network optimized for on-device inference. Unlike previous approaches reliant on specialized stethoscopes, our method directly analyzes audio recordings, facilitated by a novel architecture known as IConNet. IConNet, an Interpretable Convolutional Neural Network, harnesses insights from audio signal processing, enhancing efficiency and providing transparency in neural pattern extraction from raw waveform signals. This is a significant step towards trustworthy AI in healthcare, aiding in remote health monitoring efforts.
\end{abstract}



\maketitle


\begin{figure}[H]
    \centering
    \includegraphics[width=1\linewidth]{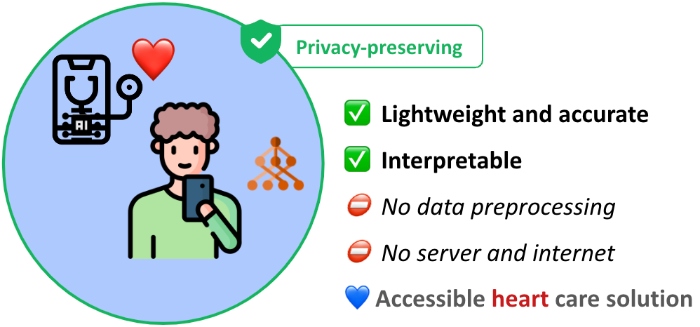}
  \caption[System illustration]{System illustration: abnormal heart sound detection using mobile phones}
  \label{fig:system}
\end{figure}

\section{Introduction}
The cardiovascular disease screening process detects abnormalities such as heart murmur, which is an irregular sound audible during the heartbeat cycle through a stethoscope. Detection of a heart murmur suggests underlying cardiac issues, prompting further evaluation through echocardiography and electrocardiography tests to pinpoint the specific heart disease. To enhance the accessibility of early diagnosis, we introduce a novel system for detecting abnormal heart sounds using mobile phones and an on-device neural network. Our system does not require extra equipment, a server, or a specific data preprocessing pipeline, which is an advantage compared to existing works. 

\section{Background}
As mobile devices become more popular and have more computational power, various systems have been developed to identify heart murmurs using mobile phones and machine learning. Ashrafuzzaman et al. \cite{ashrafuzzaman2013heart_mobile} introduced a novel system utilizing smartphone technology, including the camera and a mobile stethoscope, to estimate heart rate and detect heart attacks and other related diseases by employing Fuzzy Logic, a component of Data Mining. Thiyagaraja et al. \cite{thiyagaraja2018novel_heart_mobile} developed a mobile app that read input from a customized stethoscope to record the heartbeat sound. Their system uses both discrete and continuous wavelet transforms to downsample the audio, then extract Mel-Frequency Cepstral Coefficients (MFCC) to train a Hidden Markov Model for classification. However, these approaches have some drawbacks, such as the need for preprocessing algorithms and specific equipment. It is uncertain whether these methods would suit real-world applications where the recording environment and audio quality can vary.

As deep learning has become more prevalent with many advanced techniques to help model the data in its raw form, recent studies have been developing different deep learning models to tackle this problem. The main challenge of this approach is that long-recorded audio often includes mixed signals such as heart sound, breath, and background noise. Therefore, a complex preprocessing pipeline is mandatory. For example, Li et al.'s pipeline in \cite{li2022heart} involves a 2000Hz downsampling, a 5th-order Butterworth low-pass filtering of the 0-400Hz band and the signal pre-emphasis algorithm, before extracting MFCC and its derivatives to apply CNN or LSTM models or a combination of both. 
While achieving high accuracy, these models are complex and not mobile-friendly. In \cite{talab2019detecting_heart_mobile}, Talab et al. proposed a system that uses mobile phone recording and a deep neural network running on the server side. Sending heart sound recordings over the network to be analyzed on the server raises privacy and efficiency concerns.

\section{System design}
We propose using IConNet to identify abnormal heart sounds from mobile phone recordings. The IConNet model is an end-to-end lightweight neural network architecture that eliminates the need for heart sound segmentation, low-pass filtering, and MFCC-based feature extraction. The IConNet consists of two front-end blocks with 128 and 32 kernels, respectively, a max-pooling layer and a 2-layer feed-forward network (FFN) classifier with 256 nodes on each layer. The total number of parameters is 154180, of which 45568 parameters come from the front-end blocks. This is much smaller compared to MobileNet and MobileNet v2 models \cite{sandler2018mobilenetv2} having 4.2 million and 3.4 million parameters, respectively. The IConNet model size is 493.3 kB without further optimization, such as parameter quantization, satisfying on-device resource constraints.

\begin{figure}[ht]
    \centering
    \includegraphics[width=1\columnwidth]{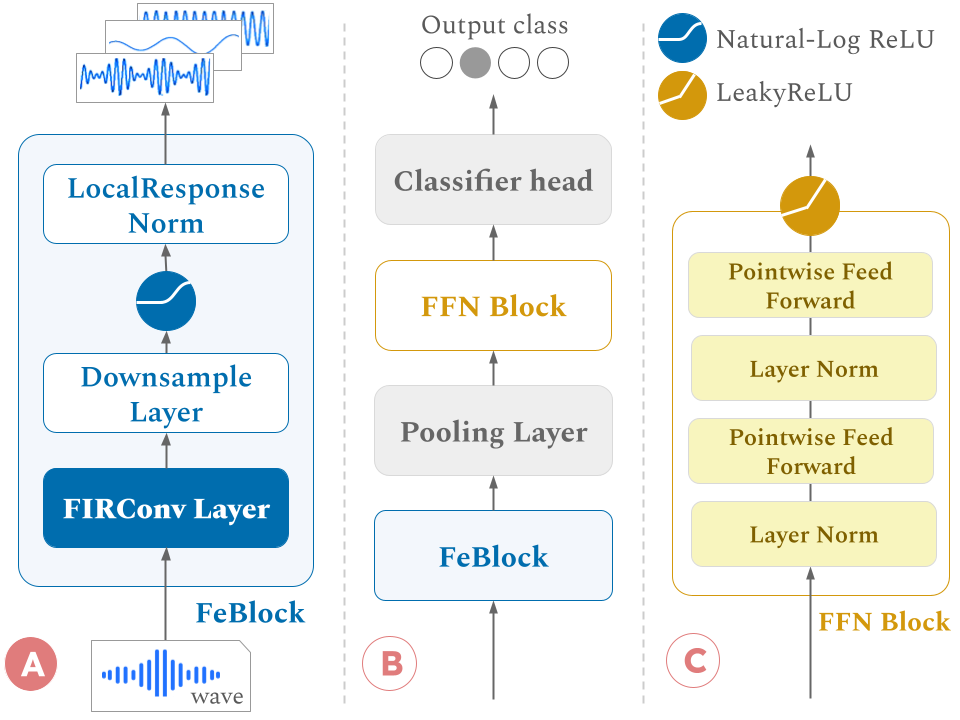}
    \caption[The IConNet architecture for end-to-end audio classification]{The IConNet architecture for end-to-end audio classification: \textit{A}- the front-end block containing the FIRConv layer; \textit{B}- the proposed general architecture for end-to-end audio classification; \textit{C}- the proposed classifier for abnormal heart sound detection.}
    \label{fig:iconnet}
\end{figure}

\section{Evaluation}
We employ the widely-used PhysioNet/CinC Challenge dataset\footnote{https://physionet.org/content/challenge-2016/1.0.0/} for heart sound classification evaluation. 
This dataset comprises 2575 \textit{normal} and 665 \textit{abnormal} samples. To validate the effectiveness of the IConNet in identifying relevant features, we resample the waveform from 2000 Hz to 16000 Hz and conduct 4-fold cross-validation, reporting UA and F1 metrics. 
Deng et al.\cite{deng2020heart} serve as our baseline, utilizing a preprocessing pipeline with a bandpass filter, MFCC features, and a CRNN model consisting of three 2D CNN layers and two LSTM layers. We also include the MFCC performance for comparison. Preprocessing for other models involves waveform trimming and downsampling, excluding the IConNet model.


\begin{table}[th]
  \caption[Abnormal heart sound detection result on the PhysioNet dataset]{Abnormal heart sound detection result on the \textbf{PhysioNet} dataset}
  \label{tab:iconnet_result_heartsound}
  \centering
  \begin{tabular}{ r@{}l c c c}
    \toprule
    \multicolumn{2}{c}{\textbf{Model}} & 
    \multicolumn{1}{c}{\textbf{UA}} & 
    \multicolumn{1}{c}{\textbf{F1}} \\
    \midrule
    MFCC + FFN & & 82.98 & 88.68 \\
    MFCC deltas + CRNN \cite{deng2020heart} & & 85.67 & 90.60 \\
    IConNet & & \textbf{87.48} & \textbf{92.05}\\
    \bottomrule
  \end{tabular}
\end{table}


\section{Discussion}
Based on the results presented in Table \ref{tab:iconnet_result_heartsound}, it is clear that the baseline model \cite{deng2020heart} performed better than the MFCC + FFN model, thanks to its preprocessing steps that included band-pass filtering and the use of MFCC deltas. The baseline model achieved a 90.06\% F1 score. However, our proposed architecture surpassed both models with an F1 score of 92.05\%, which is 2\% higher than the baseline model. While this result still does not yet outperform the state-of-the-art Resnet result reported by Li et al. in \cite{li2022heart}, it has successfully demonstrated the effectiveness of our proposed method in classifying heart sound data. 

\begin{figure}
\centering
\includegraphics[width=1\linewidth]{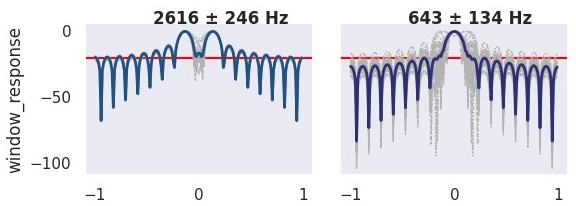}
\caption{Frequency response of filters from different bands. Each sub-figure portrays window filters of a particular frequency range, in which the average response of all corresponding filters (grey color) is presented in vibrant colored lines. The red line at -20dB represents the threshold at which noise is perceived as not noticeable.}
\label{fig:iconnet_heart_win}
\end{figure}

Furthermore, the visualization of the front-end filters in Figure \ref{fig:iconnet_heart_win} confirms that it allocates band-pass filters that actively change the window shapes to extract essential information in the range of 643 ±134 Hz. The windows have learned to transform into band-stop filter shapes for the high-frequency range above 2000 Hz, which only contain meaningless artifacts from the resampling step. Understanding the features utilized by the backbone neural network model, which influences its decisions, is vital for ensuring reliable outcomes, particularly in health applications.

\section{Conclusion}
\label{sec:conclusion}
In conclusion, this paper presents the initial phase in the development of a mobile-based heart health monitoring system. Future enhancements will enable users to track a broader range of symptoms and receive tailored recommendations for medical check-ups. Additionally, the integration of data from wearable trackers will further streamline heart health management, making it more accessible and convenient. By fostering greater awareness and proactive engagement with heart health, this system aims to empower individuals to take charge of their well-being.


\bibliographystyle{ACM-Reference-Format}

\bibliography{reference}
\end{document}